\documentstyle[epsfig,12pt,preprint,tighten,aps]{revtex}
\begin{document}
\draft
\title{
\rightline{{\tt October 2000}}
\
\\ Do ``isolated'' planetary mass objects orbit mirror stars?}
\author{R. Foot, A. Yu.\ Ignatiev and R. R. Volkas}
\address{School of Physics\\
Research Centre for High Energy Physics\\
The University of Melbourne\\
Victoria 3010 Australia \\
foot, sasha, rrv@physics.unimelb.edu.au}
\maketitle

\begin{abstract}
We propose that the ``isolated'' planetary mass objects observed by
Zapatero Osorio et al in the $\sigma$ Orionis cluster might actually
be in orbit around invisible stellar mass companions such as mirror stars.
Mirror matter is expected to exist if parity is an unbroken symmetry
of nature. 
Future observations can test
this idea by looking for a periodic Doppler shift in the radiation
emitted by the planets. The fact that the observations show an
inverse dependence between the abundance of the these objects and their mass
may argue in favour of the mirror matter hypothesis.
\end{abstract}

\newpage

A variety of 
observations strongly suggests the presence of a significant amount
of dark matter (DM) in the universe. 
Galactic rotation curves and cluster
dynamics cannot be explained using 
standard Newtonian gravity unless non-luminous
but gravitating matter exists. Arguments from Big Bang Nucleosynthesis and
theories of large scale structure formation disfavour the simple
possibility that all of the DM consists of ordinary baryons. Candidates for the
required exotic component in the DM abound: WIMPS, axions and mirror matter
are examples. The observation of microlensing events from the Small and Large
Magellanic Clouds is consistent with the existence of Massive Compact Halo
Objects (MACHOs) in the halo of the Milky Way \cite{macho}. 
The inferred average mass is about
$0.5 M_{\odot}$, where $M_{\odot}$ is the mass of our sun. The most reasonable
conventional identification 
sees MACHOs as white dwarfs, although there are several
strong arguments against this \cite{freese}. For example, the
heavy elements that would have been produced by their progenitors are not
in evidence \cite{freese}. This argues against the conventional white dwarf scenario, and
in favour of exotic compact objects. In summary, there is strong evidence for 
exotic DM which is capable of forming compact stellar mass objects.

Mirror matter \cite{mirror}
is an interesting candidate for some of the required exotic
DM \cite{blinnikov}. 
It can be independently motivated by the desire to see the full Poincar\'{e}
Group, including improper transformations (parity and time reversal), 
as an exact symmetry group of nature.
The basic postulate is that every 
ordinary particle (lepton, quark, photon, etc.)
is related by an improper 
Lorentz transformation with an opposite parity partner
(mirror lepton, mirror quark, mirror photon, etc.) of the same mass. 
Both material particles
(leptons and quarks) and force carriers (photons, gluons, $W$ and $Z$ bosons)
are doubled. Mirror matter 
interacts with itself via mirror weak, electromagnetic
and strong interactions which have the same form and strength as their ordinary
counterparts (except that mirror weak interactions couple to the opposite 
chirality).
Because ordinary matter is known to clump into compact objects such as 
stars and
planets, mirror matter will also form compact mirror stars and mirror planets.
Since mirror matter does not feel ordinary electromagnetism, it will be dark.
Gravitation, by contrast, is common to both sectors. Mirror matter therefore has
the correct qualitative features: it is dark, it clumps, and it 
gravitates.\footnote{There are several other interesting
implications of the mirror matter model. In particular,
oscillations between ordinary and mirror neutrinos have been 
proposed as a solution to the solar and atmospheric neutrino 
problems \cite{mirrornu}. Ordinary - mirror neutrino oscillations
also lead to interesting implications for early Universe
cosmology\cite{fv}. Also photon-mirror photon kinetic
mixing leads to potentially observable effects for
orthopositronium\cite{gl} and can resolve the orthopositronium
lifetime anomaly\cite{fg}.} 
Later, we will explain why the observed
inverse dependence between the abundance of the these objects and their mass
may already argue in favour of the mirror matter hypothesis.

It has been speculated that MACHOs might be mirror stars \cite{f1}, 
and one of us (RF) has proposed that the observed
extrasolar planets \cite{extrasolar} might be composed of mirror 
matter \cite{f2}. 
The former is well motivated
by the aforementioned difficulties in identifying MACHOs as conventional
stellar mass objects such as white dwarfs. The latter is motivated by the fact that
the detected extrasolar planets are rather massive and orbit very closely to the star,
which are surprising characteristics. It is unlikely that ordinary planets of sufficient
size could have condensed so close to the stars. If they are composed of ordinary matter,
then they probably formed much further from the stars and then migrated in. Another
possibility, though, is that they are composed of exotic material such as mirror matter.
Because of the weak coupling between ordinary and mirror matter, there is no barrier
to mirror planet condensation very close to an ordinary star. 
This idea should be testable
by observing the opacity and albedo properties 
of the planets\cite{foot01}. 

Zapatero Osorio et al.\ \cite{zo} have 
recently presented strong evidence for the existence
of ``isolated planetary mass objects'' in the $\sigma$ Orionis star cluster.
These objects are more massive than Jupiter $M_{J}$, 
but not as massive as brown dwarfs ($\sim 5 - 15 M_J$ although
there is some model dependence in the mass determination\cite{zo}).
They appear to be gas giant planets which do not seem to be
associated with any visible star. 
So far, eighteen such objects have been identified.
Given that the $\sigma$ Orionis cluster is estimated
to be between 1 million and 5 million years old, the formation
of these ``isolated planets'' must have occured within
this time scale.
Zapatero Osorio et al.\ argue that these findings pose a
challenge to conventional theories of planet formation
because standard theories of substellar body formation
(as well as new theories inspired by previous claims
of isolated planet discovery),
are unable to explain the existence of numerous isolated
planetary mass objects down to masses $\sim$ few $M_J$.
See Ref.\cite{zo}
and references therein for further discussion.
It is possible therefore that
non-standard particle physics may be required to understand their origin.

We speculate that rather than being isolated, these ordinary matter planets
actually orbit invisible mirror stars. These systems could be, in a sense,
just the mirror images of those ordinary star systems 
which have been speculated to
feature large Jovian mirror planets in close orbit. Indeed, if there really
are mirror planets in orbit around ordinary stars, then it is very natural to
also expect mirror solar systems to sometimes contain large ordinary planets. 

It should be possible to test this idea by searching for a periodic
Doppler shift in spectral lines emanating from these planets. We have that
\begin{equation}
\frac{\Delta \lambda}{\lambda} = 2 \frac{v_r}{c},
\end{equation}
where $\lambda$ is wavelength, $\Delta \lambda$ is the difference between
the peak and trough of the periodic Doppler modulation of $\lambda$,
$v_r$ is the maximum value of the component of the planet's orbital 
velocity in the direction
of the Earth, and $c$ is the speed of light. Suppose that a given planet
is in a circular orbit of radius $r$ around a mirror star of mass $M$. 
Let $I$ be the inclination of the plane of the orbit relative to the normal
direction defined by the Earth - mirror star line. Then
\begin{equation}
v_r = \sqrt{\frac{GM}{r}} \sin I,
\end{equation}
where $G$ is Newton's constant. Combining these equations we obtain
\begin{equation}
\frac{\Delta \lambda}{\lambda} \simeq 
10^{-3} \sqrt{\frac{M}{M_{\odot}}}
\sqrt{\frac{0.04\ A.U.}{r}} \sin I
\end{equation}
as the level of spectral resolution required. 
Note that this is a few orders of magnitude
larger than the Doppler shifts observed in extrasolar planet detection.
However it is certainly true that the isolated planets are
much fainter sources of light than the stars whose Doppler shifts
have been measured so such a measurement may not be completely 
straightforward. However, it is worth noting that for
the case of close orbiting ordinary planets
where $r \sim 0.04\ A. U.$ (analogous to the close-in extra
solar planets),
the Doppler shift is quite large ($\sim 10^{-3}$) with a 
period of only a few days which should make this
interesting region of parameter space relatively easy to test.
Indeed, Zapatero Osorio et al.\ \cite{zo} have taken optical and near infrared
low resolution spectra of three young isolated planet candidates (S Ori 52,
S Ori 56, and S Ori 47). They have obtained absorption lines
(at wavelengths $\sim 900$ nm),
however their resolution was 1.9 nm\cite{zo} which
is just below 
that needed to test our hypothesis. The higher resolution required
has been achieved in the case of brown dwarfs\cite{brown} so we
anticipate that it should be possible to test our hypothesis
in the near future. 

One would also expect some ordinary matter to have accumulated in the centre of the mirror
star. It is possible, but not inevitable, that this ordinary matter also 
observably radiates. If so,
one would expect this radiation to experience a much smaller Doppler modulation
compared to that from the planet. Because the planet and mirror star would not be 
spatially resolved, one observational signature would be that some of the spectral
lines are modulated (those from the planet), while a different set are not (those
from the ordinary matter pollutants in the mirror star). 

If the mirror star is invisible but opaque, 
then one would expect to see periodic
planetary eclipses for some of these systems (those with $\sin I \simeq 1$).
The eclipses should of course occur
once per Doppler cycle, around one of the points
of zero Doppler shift within a cycle. Obviously, such eclipses
(along with the information provided by Doppler shift measurements)
will be useful in distinguishing a mirror star from alternatives
such as faint white dwarfs or neutron stars. However, it should
be mentioned that standard objects such as white dwarfs
and neutron stars are extremely unlikely candidates, because the age
of the $\sigma$ Orionis cluser is estimated to be only 1 million to 5
million years old, while white dwarfs and neutron stars are typically
billions of years old.

Before concluding, we would like to point out an intriguing 
systematic in both
the extrasolar planet and the Zapatero Osorio et al.\ data that may 
argue in favour
of the mirror matter hypothesis. One envisages a universe 
that contained some
admixture of ordinary and mirror matter from the earliest moments after the Big 
Bang. Eventually, both the smooth ordinary fluid and the mirror fluid condensed
into large scale structures, stars and planets. Because gravitational condensation
must be aided by non-gravitational dissipative effects to carry off kinetic energy,
one does not expect the ordinary and mirror matter to have condensed in congruent locations,
despite their common gravitational interaction. One expects instead a nonzero
``segregation scale'' $\ell$ to quantify the degree of spatial separation of
condensed ordinary and mirror matter clouds or clumps. While we have
too little information to theoretically calculate $\ell$, the qualitative
expectation is a universe of cells of scale $\ell$, with a given cell being
dominated either by ordinary matter or mirror matter. Provided that $\ell$ is
much greater than a typical solar system scale, which is in fact 
observationally required,\footnote{For instance, one can deduce an upper bound
of about $10^{-3}$ for the mirror matter content of the Earth \cite{ig}.} 
then the
majority of hybrid ordinary-mirror systems should have disparate components: 
large ordinary objects with small mirror objects, or the other way around. 
The ordinary star plus mirror planet systems, and our proposed mirror star plus ordinary
planet systems, have exactly this characteristic. Indeed one might expect
the number of hybrid systems to increase as a function of the disparity between
the components. Intriguingly, the observed extrasolar planets increase in number
as their mass decreases. Even more interestingly, the Zapatero Osorio et al.\ objects
also increase in number with decreasing mass: from Fig.2 of Ref.\cite{zo} 
we see that there are
about as many objects between $8 M_{J}$ and $10 M_{J}$ as there are between
$10 M_{J}$ and $20 M_{J}$ (taking 5 million years as the relevant lifetime). 
We predict, therefore, that an extended search would
find greater numbers of these objects at even smaller masses. 
Of course if the ``isolated planets'' do orbit mirror
stars then this suggests that the star forming region
near $\sigma$ Orionis 
could also be a region of mirror star formation. This is certainly
possible and was already envisaged many years ago by 
Khlopov et al.\cite{kh} where they argued that large molecular clouds
(made of ordinary matter)
could merge with large mirror molecular clouds in which case
the formation of mixed systems (i.e. containing both ordinary
and mirror matter) is enhanced.

In conclusion, we have proposed that the ``isolated'' planetary mass objects
observed by Zapatero Osorio et al.\ might actually be planets orbiting invisible
mirror stars. This idea can be tested by searching for a Doppler modulation
at the level of $10^{-3}-10^{-4}$ in amplitude.

\acknowledgements{This work was supported by the Australian Research Council. 
R.F. is an Australian Research Fellow.}

\end{document}